\documentclass[floatfix,prb]{revtex4}

\usepackage[]{amsmath}

\usepackage[psamsfonts]{amsfonts}
\usepackage[psamsfonts]{amssymb}
	
\usepackage{graphicx}
\usepackage[usenames,dvipsnames]{color}

\newcommand{\shree}[1]{\textcolor{blue} {#1}}
\renewcommand{\shree}[1]{#1}

\newcommand{\sio} {Sr$_{2}$IrO$_{4}$ }
\newcommand{\cF} {\ensuremath{\mathcal F}}
\newcommand{\cH} {\ensuremath{\mathcal H}}

\newcommand{\threej}[6]{\ensuremath{\left(\begin{array} {ccc}#1 &#2 &#3 \\ #4 &#5 &#6 \end{array}\right)}}

\newcommand{\ninej}[9]{\ensuremath{\left\{\begin{array} {ccc}#1 &#2 &#3 \\ #4 &#5 &#6 \\ #7 &#8 &#9\end{array}\right\} }}

\def\epl{Europhys.\ Lett.}

\def\jpcm{J.\ Phys.: Condens. Matter}
\def\prl{Phys.\ Rev.\ Lett.}
\def\prb{Phys.\ Rev.\ B}

\def\aop{Ann. Phys.}

\begin{document}

\title{Non-trivial order parameter in \sio }

\author{Shreemoyee Ganguly}
\affiliation{Department of Physics and Astronomy, Uppsala University, Box 516, SE-75120 Uppsala, Sweden}
\author{Oscar Gr{\aa}n\"as}
\affiliation{Department of Physics and Astronomy, Uppsala University, Box 516, SE-75120 Uppsala, Sweden}
\author{Lars Nordstr\"om}
\affiliation{Department of Physics and Astronomy, Uppsala University, Box 516, SE-75120 Uppsala, Sweden}


\begin{abstract} 
Electronic structure calculations obtained with an approach with density functional theory with an enhanced local Coulomb interaction, DFT+$U$, are presented for the 
relativistic magnetic insulator \sio. The results are in accordance with experiments with a band gap and a small moment anti-ferromagnetic ground state. This solution is thereafter thoroughly analyzed in terms of Landau theory where it is found that the ordered magnetic moments only form a secondary order parameter while the primary order parameter is a higher order magnetic multipole of rank five. It is further observed that the electronic structure in the presence of this order parameter is related to the earlier proposed $j_\mathrm{eff}=1/2$ model, but in contrast to that model the present picture can naturally explain the small magnitude of the ordered magnetic moments.

\end{abstract}

\date{\today}


\maketitle

The recent surge in interest in 5$d$ transition metal based oxides is spurred by the combination of correlation typical for 3$d$-oxides with a much larger spin orbit coupling (SOC). Although correlations are expected to be weaker in the 5$d$ systems due to the larger extension of the 5$d$ orbitals than the 3$d$ orbitals, there have been interesting discoveries of e.g.~``relativistic Mott'' insulators \cite{jeff} as well as suggestions for relativistic topological insulators \cite{TI}. These prospective of utilizing these properties in spintronics or quantum computing applications has lead to an intense research activity on these materials \cite{Iridate-Kitaev,Kitaev-models}.
Among the Iridates \sio is regarded as the archetypical ``spin-orbit Mott" insulator, albeit with a small net magnetic moment{\cite{jeff,FYE,JN,JDAI,JKIM}}. From a theoretical view  the formation of magnetic moments in the presence of a strong spin orbit coupling, and hence without the spin as a valid quantum variable, is a very interesting and still not fully understood phenomenon.
  
  In this Letter we focus on the source of the time reversal (TR) symmetry breaking leading to the anti-ferromagnetic ground state of {\sio}. This insulating state is known  \cite{jeff,HJIN} to be well reproduced with DFT+SO+$U$ calculations (relativistic density functional theory (DFT) based calculations that include an extra local Coulomb interaction term).  
    It is found that the TR symmetry breaking cannot be described as an ordinary formation of ordered magnetic moments, but is rather best described as an ordering of higher multipoles. The observed small magnetic moments arise only as weak secondary order parameter (OP). 
The primary OP is a staggered TR odd multipole of rank five, a so-called triakontadipole. 
Its calculated stability in \sio, with its strong spin orbit coupling,  is in line with what has earlier been observed for magnetic states in actinide materials, which 
was formulated in the semi-empirical  Katt's rules \cite{Polarisation}.
 
\begin{figure}[htbp]
\begin{center}
\includegraphics[width=1.0\columnwidth]{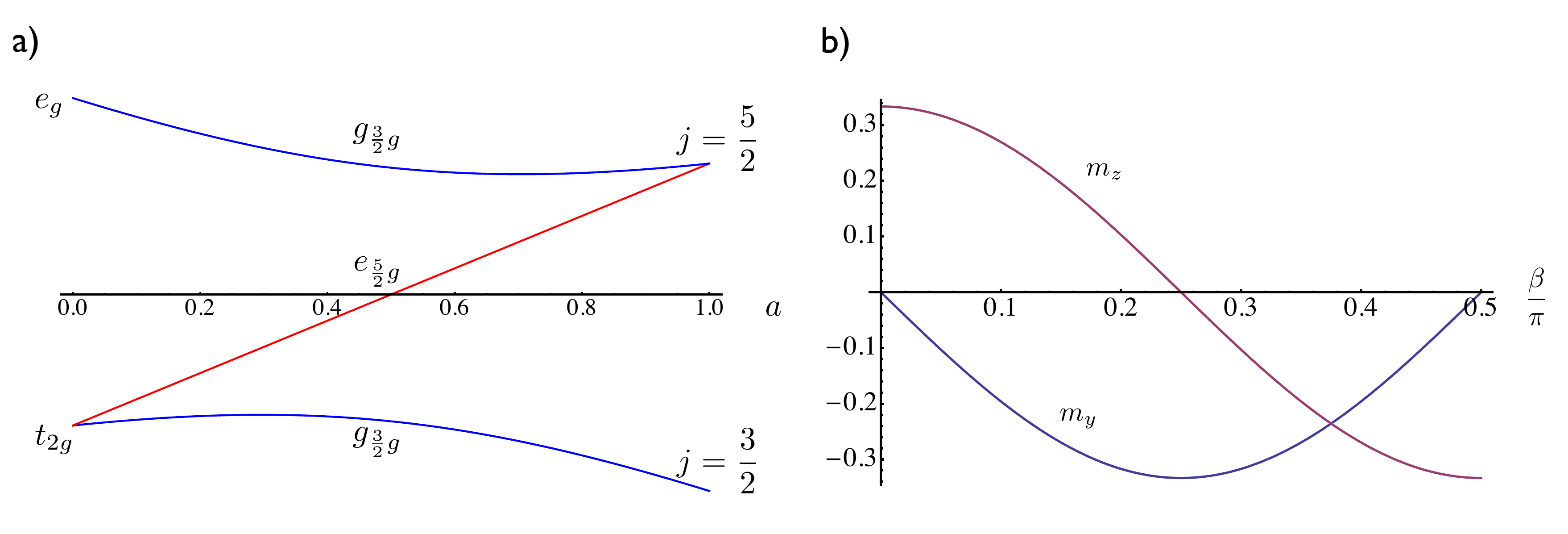}
\caption{ a) A schematic picture of the $d$ levels as an interpolation between strong spin orbit coupling $a=1$ and strong octahedral crystal field $a=0$. They belong to two different irreducible representations, $g_{3/2g}$ (blue) and $e_{5/2g}$ (red). b) The spin moments when the  state $e_{5/2g}$ is split due to a broken time reversal symmetry as a function of the parameter $\beta$ in Eq.~(\ref{beta}).}
\label{SOC+CF}
\end{center}
\end{figure}

 The combination of strong SOC, strong crystal field (CF) and strong correlation in Iridates, \sio in particular, makes construction of models inherently difficult. A model referred to as the $j_\mathrm{eff}=1/2$ model was introduced some years ago {\cite{jeff}}.
  It was based on the observation that in the presence of spin-orbit splitting in an octahedral environment, the six-fold (including spin) degenerate  states of the irreducible representation (IR) $t_{2g}$ are split into four-fold degenerate states, $g_{3/2g}$, and a Kramer doublet, $e_{5/2g}$, with labels according to the M\"ulliken notation. This splitting is schematically displayed in Fig.~\ref{SOC+CF}, where the eigenstates of $a\,\cH_\mathrm{SOC}+(1-a)\,\cH_\mathrm{CF}$  are plotted as a function of the parameter $0\leq a\leq 1$. 
  In the case of a $d$-occupation of five the Kramer doublet states $e_{5/2g}$  are half-filled and is therefore prune to split into two non-degenerate states by breaking the TR symmetry.
 
  The eigenstates of the $e_{5/2g}$ have the same form irrespective of the comparative strength of SOC and CF. In a $jm_j$-basis they take the form
\begin{align}
  \psi_1&=\cos\, \alpha \left|\frac{5}{2},\frac{3}{2}\right>-\sin\, \alpha\left|\frac{5}{2},-\frac{5}{2}\right>\nonumber\\
 \psi_2&=\cos\, \alpha \left|\frac{5}{2},-\frac{3}{2}\right>-\sin\, \alpha\left|\frac{5}{2},\frac{5}{2}\right>\,,
\label{alpha}
\end{align}
with $\alpha=\arctan\,1/\sqrt{5}$.
As any linear combination of these two states are also solutions, in general 
\begin{align}
  \psi_1(\beta,\delta)&=\cos\beta\, \psi_1 + e^{i\delta}\sin\beta\,\psi_2\nonumber\\
  \psi_2(\beta,\delta)&=\sin\beta\, \psi_1 - e^{i\delta}\cos\beta\,\psi_2\,,\label{beta}
\end{align}
describe the two-fold degenerate states. This double degeneracy is typical for the SU(2) group and that is what has given the model its name -- the $j_\mathrm{eff}=1/2$ model.
When these Kramer degenerate states are split due to TR symmetry breaking,  only the lowest state will be occupied which leads to non-vanishing expectation values of TR odd quantities such as the magnetic moments.
If only $\psi_2(\beta)$ is occupied the variation of the spin moment with the parameter $\beta$ in the case of $\delta=\pi/2$ is displayed in Fig.~\ref{SOC+CF}. It always has the magnitude $1/3$ but the moment rotates with $\beta$. The orbital moment is always parallel to the spin moment with double the magnitude.

For this model to be perfectly valid the Ir atoms should have an octahedral site symmetry with a crystal field splitting as well as spin orbit splitting much larger than the bandwidths of the $t_{2g}$ states. In reality the octahedral cages are both elongated along the $z$-axis as well as rotated in the $xy$-plane \shree{\cite{QHUANG}}, and as we will find later the band width is of a similar magnitude as the splitting due to SOC.
Experiments both in favor and disfavor of the model exist. Therefore there is a vivid debate in recent literature about the applicability of this model both for 
the case of iridates in general and for \sio in particular. \shree{\cite{RARITA,GCAO,HWATANABE,JKIM,FYE,HJIN}} 

 A clear short come of this model is that it cannot explain the much smaller magnetic ordered moments found on the Ir sites, around 0.3 $\mu_\mathrm{B}$ instead of 1 $\mu_\mathrm{B}$, instead it is hand-wavingly argued that the strong reduction stem from strong hybridizations with ligand states. \cite{jeff}

In order to investigate the details of the physics a realistic electronic structure is obtained with the APW+{\em lo} method in the DFT+SO+$U$ approach, as implemented in the code {\sc Elk} \cite{LAPW,Bultmark-Mult,elk}.
A magnetic unit cell in line with experimental observations \cite{JKIM,FYE,QHUANG} with eight formula units were used in all calculations. 
The $U$ parameter was expressed as a linear combination of Slater parameters which in turn was calculated as radial integrals of the Yukawa screened Coulomb potential and the localized limit was adopted for the double counting correction. \cite{Bultmark-Mult}

%
 
\begin{figure}[htbp]
\begin{center}
\includegraphics[width=0.8\columnwidth]{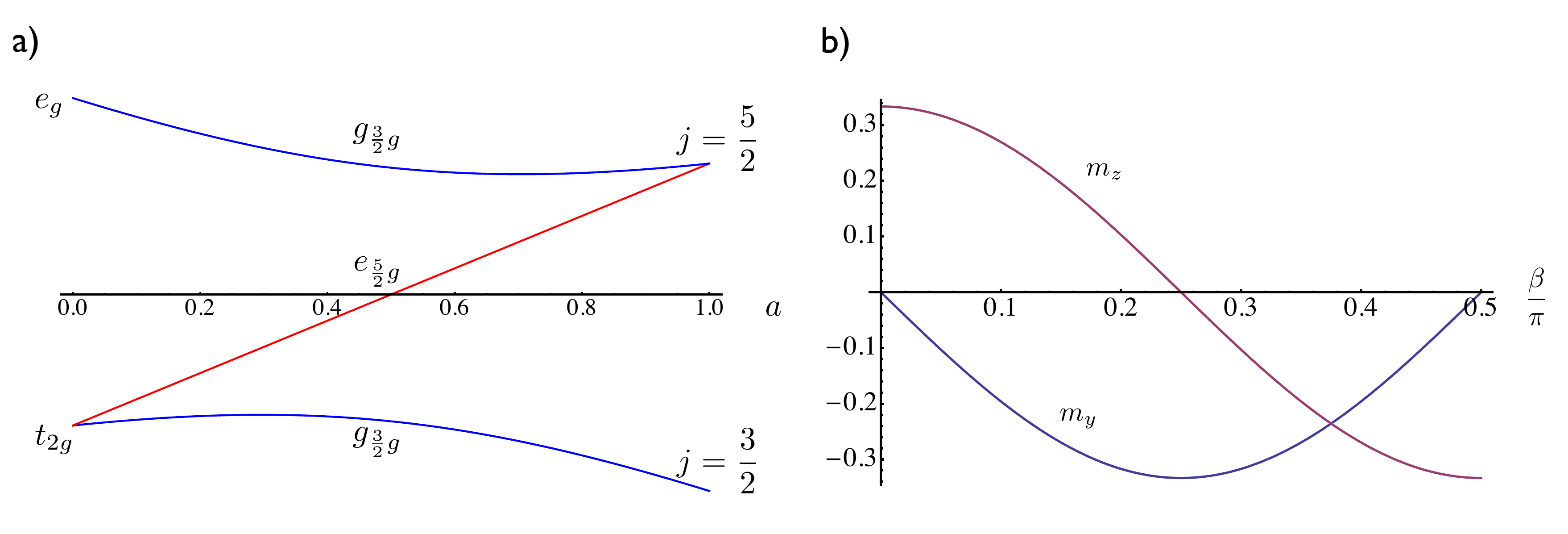}
\caption{Comparison between the energies of the TR even and the TR odd (AF) solutions $\Delta E$ and the magnitude of the energy gap $E_\mathrm{gap}$ in the TR odd case as a function of the parameter $U$ in the DFT+SO+$U$ method. Note that $U=0$ eV corresponds to the DFT-limit.
      \label{momU}}
\end{center}
\end{figure}

 Several calculations allowing for a magnetic solution with varying values of the Hubbard-$U$ parameter have been performed, the results of which are displayed in Fig.~\ref{momU}. Here the energy difference $\Delta E$ between the TR odd and TR even solution is plotted together with the magnitude of the band gap $E_\mathrm{gap}$ in the former case.

 In the pure DFT limit, $U=0$, the TR odd anti-ferromagnetic (AF) solution is only meta stable and of metallic character. 
 With increasing $U$ the AF solution first  becomes insulating for a value of $U$ just below 2 eV  and then this solution also becomes stable for a value just above 2 eV. For a value of $U=3$  eV the result is in good accordance with experiments \cite{FYE,JN,JDAI,JKIM} in many respects 
as well as with earlier similar calculations \cite{jeff,HJIN}.

It has magnetic moments consisting of a spin part of 0.08 $\mu_\mathrm{B}$ and an orbital part of 0.24 $\mu_\mathrm{B}$ per Ir atom along the $b$-axis, with smaller components along the $a$-axis. The experimental values for the total local moment varies between 0.21 and 0.36 $\mu_\mathrm{B}$, which compares well with our calculated value of 0.32 $\mu_\mathrm{B}$. The Ir local moments are arranged in the anti-ferromagnetic order that was given in Ref.~\onlinecite{FYE}. All the calculated Ir local moments have same magnitude with an angle of $9^{\circ}$ off the $b$-axis, which is close to the rotational angle of the Ir centered oxygen octahedra and in good accordance with experimental estimates.

\begin{figure}[htbp]
\begin{center}
\includegraphics[width=0.9\columnwidth]{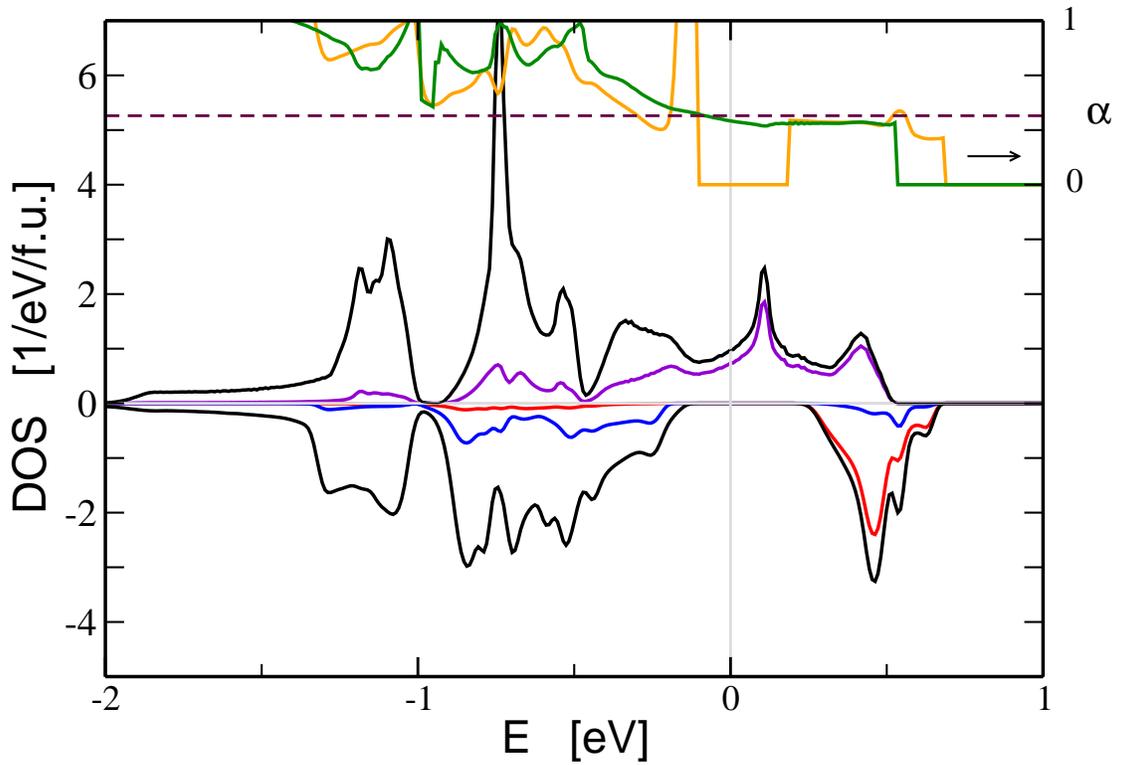}
\caption{ The Ir-$d$ projected density of states (black) for the TR even solution (upper positive part) and TR broken solution (lower negative part). Also the $j=\frac{3}{2}$, $m_j=\pm3$ (violet) projection in the case of TR even 
and the $\phi_1$ (red) and $\phi_2$ (blue) projections in case of TR odd [refer Eq.~(\ref{eigvec})] solutions are displayed. In addition an estimate of $\alpha$ of Eq.~(\ref{alpha}) is compared to the ideal $e_{5/2g}$ value for both the TR even (green) and the TR odd (orange) cases in the inset.     \label{dos}}
\end{center}
\end{figure}
In Fig.~\ref{dos} upper panel  the total  Ir-$d$ projected density of states (DOS) is displayed, together with a projection on the states with  $j=5/2$ and $m_{j}=\pm 3/2$ character, for the TR even solution in the case of $U=3$ eV. For the main energy window the displayed states correspond to the ``$t_{2g}$"-like bands. 
As discussed above, in the presence of spin orbit coupling these six degenerate states would split into the four-fold degenerate $g_{5/2g}$ IR and the doubly degenerate $e_{5/2g}$ IR.

 An inset also shows an estimate of $\alpha$ of Eq.~(\ref{alpha}) as $\arctan\sqrt{D_{5/2,-5/2}/D_{5/2,3/2}}$ (where $D_{5/2,-5/2}$ and 
$D_{5/2,3/2}$ are the DOS for $j=5/2$; $m_{j}=-5/2$ and  $m_{j}=3/2$ respectively) together with the ideal value expected in the IR $e_{5/2g}$ in the case of octahedral symmetry. We see from Fig.~\ref{dos} (upper panel) that the 
$j=\frac{5}{2}$, $m_j=\pm3$ dominate the Ir-projected states near the Fermi level. This together with the fact that $\alpha$ is 
very close to the ideal value for a large range in energy, from just below the Fermi energy to the top of the ``$t_{2g}$"-band, we can conclude that the corresponding TR symmetric Ir-projected DOS is very well described as pure $e_{5/2g}$ states.

\begin{widetext}
Now in order to further analyze the magnetic solution stabilized with $U=3$ eV, we have performed calculations where the major component of the staggered local spin moments $m_{y}(\vec{R}_n)$ are constrained \cite{Dederichs} by auxiliary constraining fields $h_{s,y}(\vec{R}_n)$,
where $n$ runs over the Ir atoms (with volume $S_n$),
\begin{align} 
E\left(m_{y}\right)=\min \left\{ E_{DFT+U} + \sum_n h_{s,y}(\vec{R}_n) \left(\int_{S_n} \hat{y}\cdot \vec{m}(\vec{r})\mathrm{d}V - m_{y}(\vec{R}_n)\right)\right\}\,.\label{fixspin}
\end{align}
\end{widetext}

\begin{figure}[htbp]
\begin{center}
\includegraphics[width=1.0\columnwidth]{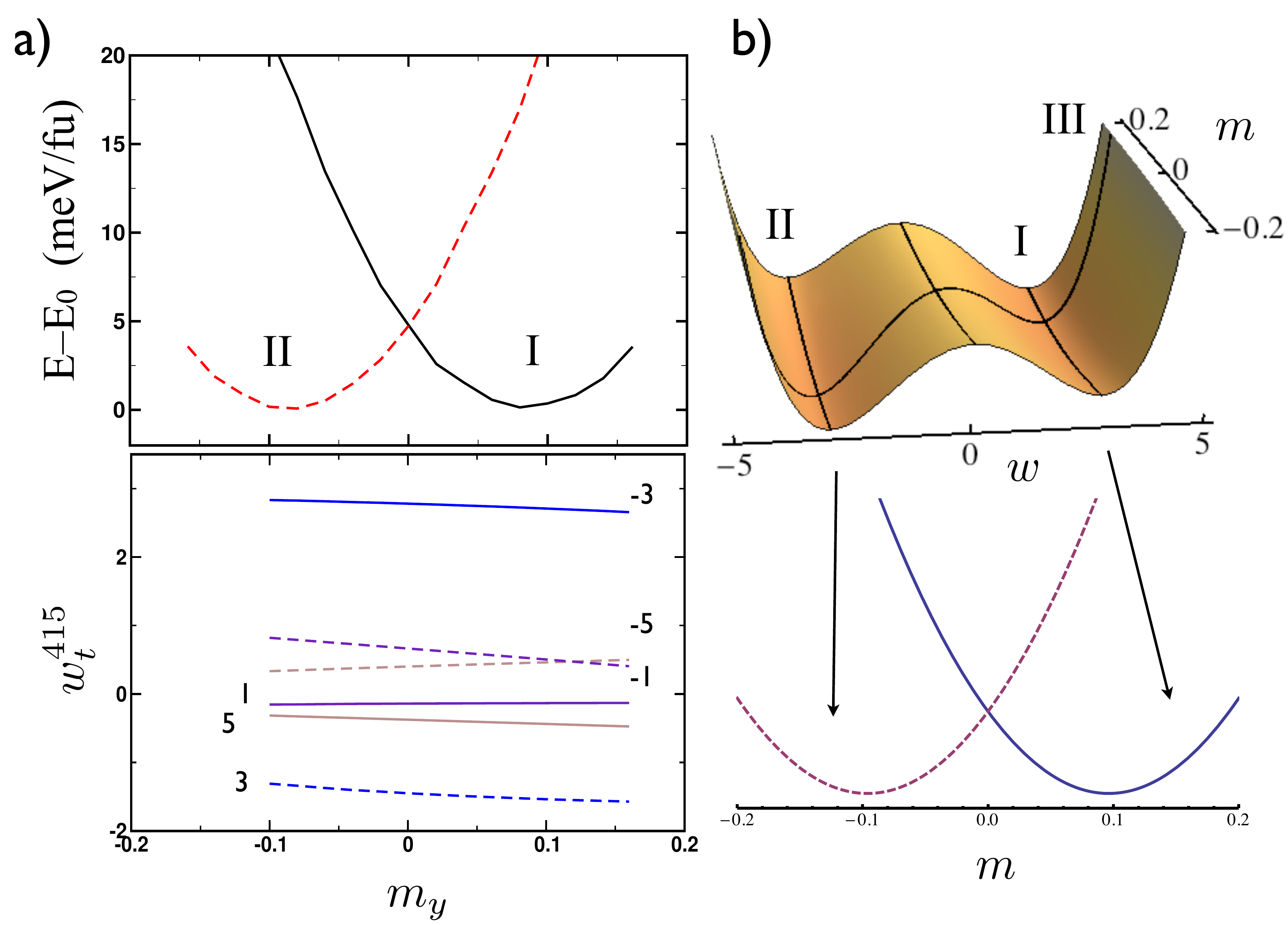}
 \caption{ a) The total energy as a function of the constraining spin magnetic moment component $m_y$ at the Ir site, for two degenerate solutions {\sc I} and {\sc II}, respectively (upper panel), together with the largest tensor components of the triakontadipole $w^{415}$ for the case {\sc I} (lower panel). b) A schematic plot of the
free energy expression of Eq.~\eqref{landau} together with the constrained moment curves, {\sc I} and {\sc II}, and the constrained primary order parameter curve {\sc III}.
      \label{Emom}}
\end{center}
\end{figure}

The variation of the energy is calculated around the equilibrium solution as displayed in curve I of Fig.~\ref{Emom}a. The energy shows a simple quadratic behavior as it should. However, there is nothing special about the $m_{y}=0$ value --- the energy does not posses a mirror symmetry for $m_{y}\rightarrow -m_{y}$ as is expected for a TR odd OP. 
Instead if we perform a similar constrained calculation starting around a solution with all magnetic moments switched, we get the curve indicated by II in Fig.~\ref{Emom}a, which is the TR mirror of curve I.
This is a strong indication that the spin moment is only a secondary OP induced by the ordering of a yet undetermined primary OP.
%
 
 To illustrate this we compare our calculated results with the most simple model for 
 two independent TR odd OP $w$ and $m$ that interact weakly with each other. The corresponding Landau free energy $\cF[m,w]$ is
\begin{align} 
\cF[m,w]=
a_{0}m^{2}+b_{0}w^{2}+b_{1}w^{4}+gm\cdot w + \dots \,,\label{landau}
\end{align}
where terms such as $a_{1}m^{4}$ and higher terms can be omitted.

Below a certain temperature, where the coefficient $b_{0}\leq 0$ the primary OP $w$
 spontaneously gets a value $w_{0}\neq 0$, determined by $\frac{\partial\cF}{\partial w}=\frac{\partial\cF}{\partial m}=0$.
At this energy minimum the secondary OP $m$ gets an induced value through the interaction term although the coefficient $a_{0}>0$.
This induced moment becomes 
$m_{0}= -{g} w_{0}/{2a_{0}}$.
One can obtain the energy variation with $m$, $\cF(m)$,  around this minimum when $w$ is simultaneously optimized, i.e.~under the constrain $\frac{\partial\cF}{\partial w}=0$. 
For a weak interaction the primary OP is essentially constant $w\approx w_{0}$ and we have that
\begin{align} 
\cF(m) =a_{0}m^{2}+gm\cdot w\approx a_{0}(m-m_{0})^{2}+\mathrm{const}\,.\label{fixspin2}
\end{align}
 There exist two degenerate energy minima at $\{m_{0},w_{0}\}$ and $\{-m_{0},-w_{0}\}$, since the free energy has to be TR symmetric, $\cF[m,w]=\cF[-m,-w]$. Hence one can deduce that there have to be two independent parabolas of Eq.~(\ref{fixspin2}) centered at $\pm m_{0}$, respectively.

The free energy variation of Eq.~(\ref{fixspin2}) closely corresponds to the fixed $y$-spin component calculations of Eq.~(\ref{fixspin}) displayed in Fig.~\ref{Emom}a. This is illustrated by the schematic free energy landscape plot of Fig.~\ref{Emom}b with appropriate values of the coefficients $a_{0}$, $b_{0}$, $b_{1}$ and $g$. In the latter plot the contours indicate $\frac{\partial\cF}{\partial m}=0$ and $\frac{\partial\cF}{\partial w}=0$, respectively. Two of the contours that pass through either of the two minima, I and II, correspond to 
the $\cF(m)$ curves while the third contour, III, is defined by  $\frac{\partial\cF}{\partial m}=0$ and  goes through both minima as well as the point $w=m=0$. The former two curves are not individually mirror symmetric in $m$ while the latter is mirror symmetric in $w$. This summarize the criteria which points out $m$ as a secondary OP induced by an underlying primary OP $w$.


Hence, the result with a shifted parabola centered around the equilibrium local moment in Fig.~\ref{Emom}a is a clear sign that the local moment is only an induced secondary OP. This leaves the question  which is the primary OP? In order to answer this question we follow the general multipole analysis \cite{Bultmark-Mult} 
that has recently been developed. This nalysis is especially powerful for 
analyses of DFT+$U$ calculations \cite{Polarisation} and  technical details relevant for the present study are given as Supplementary Materials \cite{SI}.
The expectation value of multipole tensors on Ir site $n$ are defined through 
\begin{align} 
{w}^{kpr}_{t}(n)=
 \left< d^{\dagger}_{n}\, \Gamma^{kpr}_{t} d^{\phantom{\dagger}}_{n}\right>\ ,
\label{eq:w_kpr}
\end{align}
where in the ten-dimensional space of local $d$-orbitals, $\Gamma^{kpr}_{t}$ is an hermitian matrix-operator \cite{Polarisation} and the creation operator $d^{\phantom{\dagger}}_{n}$ is a vector-operator. 
For a $d$-shell $0\le k\le 4$, $0\le p \le 1$ and $|k-p|\le r\le k+p$, which altogether constitute 18 different multipole tensors with a total number of tensor components of 100. These then fully accounts for the freedom of the ten-dimensional density matrix $\left<d^{\phantom{\dagger}}_{n}d^{\dagger}_{n} \right>$. 


In the lower panel of Fig.~\ref{Emom}a the components of the multipole with largest polarization  as well as largest contribution to the intra-atomic exchange energy \cite{Polarisation} are displayed; the triakontadipole (rank 5) tensor $w^{415}$. All odd components are non-zero which is intimately connected with the in-plane anisotropy. Here it is worthwhile to observe that the largest components $w^{415}_{\pm3}$ are almost constant under variation of $m_{y}$. This fact together with the fact that they have the largest contribution to the exchange energy, 70 meV/atom to be compared to 2 meV/atom for the spin polarisation, single them out as the primary OP of the time reversal broken symmetry solution of \sio.

The formation of this OP can be understood from the following. 
The linear combination of tensor components $w^{415}_{\pm3}$ can be viewed as a rotation of the largest component, $w^{415}_{-3}$  by an angle $\theta$ around the $z$-axis. Then for the corresponding operator
\begin{align} 
\tilde{\Gamma}^{415}_{-3}
&=
e^{-i\theta J_{z}}{\Gamma}^{415}_{-3}e^{i\theta J_{z}}=\cos 3\theta\, \Gamma^{415}_{-3}-\sin 3\theta\, \Gamma^{415}_{3}\label{rot}
\end{align}
The two largest eigenvalues (in magnitude) of this rotated operator are $\lambda_{1,2}=\pm\sqrt{70}$  and the corresponding eigenvectors are
\begin{align} 
 \phi_1&=\frac{1}{\sqrt{2}}\left\{ \left|\frac{5}{2},\frac{3}{2}\right>+i e^{3i\theta}\left|\frac{5}{2},-\frac{3}{2}\right>\right\}\nonumber\\
\phi_2&=\frac{1}{\sqrt{2}}\left\{ \left|\frac{5}{2},\frac{3}{2}\right>-i e^{3i\theta}\left|\frac{5}{2},-\frac{3}{2}\right>\right\}\,,\label{eigvec}
\end{align}
respectively.
The appearance of TR odd multipole tensors in the ground state gives rise to splitting of the TR even solution by an auxiliary field, which is  a matrix in the local basis and proportional to the magnitude of the rotated tensor moment $\tilde{w}^{415}_{-3}$  
\begin{align}
2  K_{415}\, \tilde{w}^{415}_{-3}\, \tilde{\Gamma}^{415}_{-3}(\theta)\,,\label{field}
\end{align}
where $K_{415}$ is  a known  linear combination of the three Slater (or Racah) parameters. \cite{Polarisation}
Now we can readily see that the presence of the OP $\tilde{w}^{415}_{-3}$ primarily splits the degenerate $j=5/2$, $m_j=\pm3/2$ 
states, that  dominate around the Fermi energy for the TR even case, through the action of Eq.~\eqref{field}. 
This is illustrated by the DOS projected upon these eigen-vectors $\phi_1$ and $\phi_2$ that are shown in the lower panel of 
Fig.~\ref{dos}. The splitting is almost complete with only a tiny occupation of the $\phi_1$ states while the $\phi_2$ states are almost fully occupied. 
However, after the TR breaking the states are not predominantly $j=5/2$, $m_j=\pm3/2$ anymore, which is most clearly seen for the occupied states in Fig.~\ref{dos}. 
Here the $\phi_2$ are strongly hybridizing with the other Ir-$d$ states, with which they now overlap in energy.

It is noteworthy to observe that these states are related to the TR odd states of the $j_\mathrm{eff}=1/2$ model 
of Eq.~(\ref{alpha}) -- in fact $\phi_i\equiv\psi_i(\pi/4,\pi/2+3\theta)$ when 
$\alpha=0$. However, for the states in the TR odd case we observe from the estimate of the angle $\alpha$ in the inset of Fig.~\ref{dos} that the $\phi_2$ states deviate appreciably from the ones of the $j_\mathrm{eff}=1/2$ model.

Finally we observe that the multipole rotation angle, that can be obtained from the expectation value of the tensor moments $w^{415}_{\pm3}$ of Fig.~\ref{Emom}a through Eq.~\eqref{rot}, 
is around $9^{\circ}$ which
is very close to the surrounding octahedron rotation and do not vary much with the constraining $m_y$. This is in accordance with the recent experimental observation \cite{BWVS} that the spin moment is coupled to the octahedral rotations. However, in the light of our calculations this coupling is due to a two step process, the spin moment as a secondary OP  is coupled to the primary OP, the triakontadipole, which in turn is coupled to the oxygen octahedra.


The support  from 
the Swedish Research Council (VR) is thankfully acknowledged. 
The calculations have 
been performed at the Swedish high performance centers 
HPC2N and NSC 
under grants provided by the 
Swedish National Infrastructure for Computing (SNIC).

\newpage

{\Large Supplementary Materials}
\bigskip
 
In this study we are interested in a combination of
strong spin orbit coupling (SOC), strong crystal field (CF) and significant correlation ($U$).
These effects are included in the very simple model
\begin{align} 
\cH=\cH_\mathrm{SOC}+\cH_\mathrm{CF}+\cH_{U}\,,\label{ham}
\end{align}
and are  manifested in for instance \sio which has Ir atoms positioned in quasi-octahedral sites. 
In the main paper we present results obtained by an all-electron full-potential electronic structure calculation, within the APW+{\em lo} method. Here we now present details of the tools used in the subsequent analysis of the obtained results and for this purpose the simple model Hamiltonian of Eq.~\eqref{ham} suffices. 

\subsection{CF and SOC and the $j_\mathrm{eff}$ model} 
\label{CF-SOC}
In the simple Hamiltonian of Eq.~\eqref{ham}, the term $\cH_\mathrm{SOC}$ is diagonal in a $j m_j$ basis (ordered with increasing $m_j$ for each $j=\{3/2,5/2\}$) while the term $\cH_\mathrm{CF}'$  is diagonal in a  tesseral $m_\ell m_s$ basis (which we here order as $xy,yz,xz,z^2,x^2-y^2$ for each spin component),
\begin{align} 
\cH_\mathrm{SOC}&=\xi\,\ell\cdot s=\frac{\xi}{2}\mathrm{diag}\{{-3,-3,-3,-3,2,2,2,2,2,2}\}\\
\cH_\mathrm{CF}'&=D\,\mathrm{diag}\{-2, -2, -2, 3, 3, -2, -2, -2, 3, 3\}\,,
\end{align}
while $\cH_U$ is of more complicated form and will be treated in a mean-field fashion below.

For simplicity we neglect the last term to start with.
Now we prefer to work in the $j m_j$ basis, which at first might look unusual but we will later find it to be the easiest choice. 
Hence we need to transform
$\cH_\mathrm{CF}=V^{\phantom{\dagger}} \cH_\mathrm{CF}' V^{\dagger}$, with the basis transformation $V$ that is a combination of three independent transformations: 
\begin{enumerate}
\item
 from a spherical $m_\ell m_s$  to $j m_j$ basis given by Clebsch-Gordan coefficients, or equivalently as here, Wigner 3j coefficients (with $\ell=2$ and $s=1/2$): 
\begin{align} 
V^\mathrm{CG}_{j,m_j;m_{\ell},m_s}=\sqrt{2j+1}(-1)^{\ell-s-m_j}\threej{j}{\ell}{s}{-m_j}{m_\ell}{m_s}
\end{align}
\item a tesseral $m_{\ell}'$ to spherical (with Condon-Shortley phase convention) $m_\ell$ harmonics transformation
\begin{align} 
V^\mathrm{TS}_{m_\ell,m_s;m_{\ell}',m_s'}=\delta_{m_s,m_{s'}}\left\{
\begin{array}{lr}
(-1)^{m_\ell}\left[\delta_{m_\ell,m_{\ell}'}+ i\delta_{m_\ell,-m_{\ell}'}\right]/\sqrt{2};&m_\ell>0\\
\delta_{m_\ell,m_\ell'};&m_\ell=0\\
(-1)^{m_\ell}\left[\delta_{m_\ell,-m_{\ell}'}-i\delta_{m_\ell,m_{\ell}'}\right]/\sqrt{2};&m_\ell <0
\end{array}\right.
\end{align}
\item a reordering of the tesseral components from the most natural, $\{xy,yz,z^2,xz,x^2-y^2\}$, to an order more suitable to the splitting of the $t_{2g},e_g$ IR, $\{xy,yz,xz,z^2,x^2-y^2\}$, i.e.~by exchanging third and fourth row through $V^{34}$.
\end{enumerate}
Then we have that
\begin{align} 
V&=V^\mathrm{CG}\left(V^\mathrm{TS}\otimes 1\right)\left( V^{34}\otimes 1\right)=\\
&\left(
\begin{array}{cccccccccc}
 i \sqrt{\frac{2}{5}} & 0 & 0 & 0 & 0 & -\frac{i}{\sqrt{10}} & 0 & 0 & 0 & \frac{i}{\sqrt{2}} \\
 0 & i \sqrt{\frac{3}{10}} & 0 & \frac{i}{\sqrt{10}} & 0 & 0 & -\frac{i}{\sqrt{5}} & 0 & -i \sqrt{\frac{2}{5}} & 0 \\
 0 & \sqrt{\frac{3}{10}} & 0 & -\frac{1}{\sqrt{10}} & 0 & 0 & -\frac{1}{\sqrt{5}} & 0 & \sqrt{\frac{2}{5}} & 0 \\
 0 & 0 & \sqrt{\frac{2}{5}} & 0 & 0 & 0 & 0 & -\sqrt{\frac{3}{5}} & 0 & 0 \\
 \sqrt{\frac{2}{5}} & 0 & 0 & 0 & 0 & -\frac{1}{\sqrt{10}} & 0 & 0 & 0 & -\frac{1}{\sqrt{2}} \\
 0 & 0 & 0 & i \sqrt{\frac{2}{5}} & -\frac{i}{\sqrt{2}} & 0 & 0 & 0 & \frac{i}{\sqrt{10}} & 0 \\
 -\frac{i}{\sqrt{10}} & 0 & -i \sqrt{\frac{3}{10}} & 0 & 0 & -i \sqrt{\frac{2}{5}} & 0 & -\frac{i}{\sqrt{5}} & 0 & 0 \\
 -\frac{1}{\sqrt{10}} & 0 & \sqrt{\frac{3}{10}} & 0 & 0 & -\sqrt{\frac{2}{5}} & 0 & \frac{1}{\sqrt{5}} & 0 & 0 \\
 0 & -\sqrt{\frac{2}{5}} & 0 & 0 & 0 & 0 & -\sqrt{\frac{3}{5}} & 0 & 0 & 0 \\
 0 & 0 & 0 & -\sqrt{\frac{2}{5}} & -\frac{1}{\sqrt{2}} & 0 & 0 & 0 & -\frac{1}{\sqrt{10}} & 0 \\
\end{array}
\right)\,,
\end{align}
where $\otimes 1$ indicate the block diagonal form in the spin indices.

The result of the eigenvalue problem
\begin{align} 
\left\{a\cH_\mathrm{SOC}+(1-a)\cH_\mathrm{CF}\right\}\psi=\varepsilon\psi\,,\label{phi-deg}
\end{align}
for the case $\xi/4=D=1$, was displayed in Fig.~1 of the main paper. There are three different eigen-states, belonging to the irreducible representations  (IR) $g_{3/2g}$ and $e_{5/2g}$ (in Mulliken notation). This result is true for any value of $0<a<1$. For a $d$ occupation around $n_d=5$ it is the doubly degenerate middle state $e_{5/2g}$ that is close to half-filled.
This is the essential ingredients of the $j_\mathrm{eff}=1/2$ model that was introduced for iridates ($n_d\approx5$) some years ago {\cite{jeff}}.

  The eigenstates of the $e_{5/2g}$ have the same form irrespective of the comparative strength of SOC and CF. In a $jm_j$-basis they take the simple form
\begin{align}
  \psi_{1}^T&=\left(0,0,0,0,0,\sqrt{5},0,0,0,-1\right)/\sqrt{6}\\
 \psi_{2}^T&=\left(0,0,0,0,-1,0,0,0,\sqrt{5},0\right)/\sqrt{6}\,.
\end{align}
Any linear combination of these degenerate states are also solutions, in general 
\begin{align}
  \psi_1(\beta,\delta)&=\cos\beta\, \psi_1 + e^{i\delta}\sin\beta\,\psi_2\nonumber\\
  \psi_2(\beta,\delta)&=\sin\beta\, \psi_1 - e^{i\delta}\cos\beta\,\psi_2\,,\label{beta}
\end{align}
describe the two-fold degenerate states. 
When these Kramer degenerate states are split due to TR symmetry breaking,  only the lowest state will be occupied which leads to non-vanishing expectation values of TR odd quantities such as the magnetic moments as illustrated in Fig.~1 of the main paper.

\subsection{The correlation term} 
\label{CORR}
The TR symmetry breaking have to come from the third term $\cH_U$ in the simple Hamiltonian of Eq.~\eqref{ham}.  Starting from a rotational invariant local Coulomb interaction, which is essential for cases with strong SOC,  and treating it in the mean field limit it has been shown  \cite{Bultmark-Mult,Polarisation} that this term can be expanded in multipole tensors. Since we are mainly interested in TR odd contributions we can concentrate on the exchange part of the Coulomb interaction. Due to correlations this is statically screened and we refer to it as the screened exchange interaction ($E_\mathrm{X}$),
\begin{align} 
E_\mathrm{X}(n)=\sum_{kpr} E_\mathrm{X}^{kpr}(n)=\sum_{kpr;t} K_{kpr} {w^{kpr}_t(n)}^2\,.
\end{align}
This results in an effective one-body Hamiltonian of the form
\begin{align} 
\cH_U(n)=\frac{\partial E_\mathrm{X}(n)}{\partial \rho_n^T}=
2\sum_{kpt;t}K_{kpr}\,{w^{kpr}_t}(n)\, \frac{\partial}{\partial \rho_n^T}\,{w^{kpr}_t}(n)=
2\sum_{kpt;t} K_{kpr}\, w^{kpr}_t(n)\, \Gamma^{kpr}_t\,,\label{hamu}
\end{align}
where $w^{kpr}_t$ are expectation values for the tesseral component $t$ of the multipole tensor $w^{kpr}$, $\Gamma^{kpr}_t$ are the corresponding tensor operators and $K_{kpr}$ is an energy parameter which is a linear combinations of the Slater (or Racah) parameters that describe the local Coulomb interaction.


The derivative in Eq.~\eqref{hamu} follows directly from the following definition of $w^{kpr}_t(n)$.
The expectation value of multipole tensors on Ir site $n$ are defined through 
\begin{align} 
{w}^{kpr}_{t}(n)=\mathrm{Tr} \, {\Gamma}^{kpr}_{t}\rho_{n} \,=
\mathrm{Tr} \,{\Gamma}^{kpr}_{t} \left< d^{\phantom{\dagger}}_{n}d^{\dagger}_{n} \right>=
\mathrm{Tr} \left<{\Gamma}^{kpr}_{t}  d^{\phantom{\dagger}}_{n}d^{\dagger}_{n} \right>=
 \left< d^{\dagger}_{n}\, \Gamma^{kpr}_{t} d^{\phantom{\dagger}}_{n}\right>\ ,
\label{eq:w_kpr}
\end{align}
where in the ten-dimensional space of local $d$-orbitals, $\Gamma^{kpr}_{t}$ is an hermitian matrix-operator \cite{Polarisation} and the creation operator $d^{\phantom{\dagger}}_{n}$ is a vector-operator. 

For a $d$-shell the multipole tensor moments are enumerated through the variations $0\le k\le 4$, $0\le p \le 1$ and $|k-p|\le r\le k+p$, which altogether constitute 18 different multipole tensors with a total number of tensor components ($-r\leq t\leq r$) of 100. These then fully accounts for the freedom of the ten-dimensional density matrix $\left<d^{\phantom{\dagger}}_{n}d^{\dagger}_{n} \right>$. 

In a matrix-representation in the $jm_{j}$-basis of the $d$-states 
 the multipole tensor operators are expressed in terms of Wigner $3j$ and $9j$ operators as
\begin{align}
{\Gamma}^{kpr}_{t;(j_1m_{j1})(j_2m_{j2})}=&
\frac{\sqrt{(2j_{1}+2)(2j_{2}+1)}} {N_{kpr\ell}}(-)^{k+p+r}
 \ninej{\ell} {\ell} {k} {s} {s} {p} {j_{1}} {j_{2}} {r} \, 
(-)^{j_{1}-m_{1}}
\mathcal{T}  \threej{j_{1}} {r} {j_{2}} {-m_{1}} {t} {m_{2}}
\label{Gamma}\,,
 \end{align}
with $N$ being a normalization factor \cite{Bultmark-Mult} and
where the operator $\mathcal{T}$ brings a spherical tensor to a tesseral form,  \begin{align} 
\mathcal{T} a_{t}=
\left\{\begin{array} {lr} 
\sqrt{2}\, (-1)^t\, \Re\, a_{t}& t> 0\\
{a}_{t}& t=0\\
  -\sqrt{2}\,(-1)^t\, \Im\, a_{t} 
  & t< 0\\
  \end{array}\right.\,,
\end{align}
which in turn ensures the hermitean property of $\Gamma$.

The energy parameter $K_{kpr}$ are the same as in Ref.~\onlinecite{Bultmark-Mult} and related to the coefficients $A_k$ of Ref.~\onlinecite{Polarisation} through
\begin{align} 
K_{kpr}=-\frac{(2k+1)(2p+1)(2r+1)}{2}|N_{kpr\ell}|^2\,(2\ell+1)\,A_k\label{K_kpr}
\end{align}
In the latter paper it is clear that they can be expressed in Racah parameters through
\begin{align} 
A_k=\sum_{i=0}^\ell C^{(\ell)}_{ki} E^{(i)}\,,\label{A_k}
\end{align}
where the coefficients $C^{(\ell)}_{ki}$ are explicitly given for $\ell \leq 3$. 
Noticeable is that all the coefficients $\tilde{C}^{(\ell)}_{ki}=(2\ell+1)\,C^{(\ell)}_{ki}$  are integers.
In this study we adapt the most common normalization convention \cite{vdL,Bultmark-Mult} 
\begin{align} 
N_{kpr\ell}=
i^{k+p+r} \left[\frac{(g-2k)!(g-2p)!(g-2r)!}{(g+1)!}\right]^{1/2}\frac{g!!}{(g-2k)!!(g-2p)!!(g-2r)!!}\,n_{lk}\,n_{sp} \ ,
\end{align}
where $g=k+p+r$, and 
\begin{align} 
n_{\ell k}&=\frac{(2\ell)!}{\sqrt{(2\ell-k)!(2\ell+k+1)!}}
\end{align}
which means that for $s=1/2$ and $p\in\{0,1\}$
\begin{align} 
n_{sp}=\frac{1}{\sqrt{2(2p+1)}}\,.
\end{align}

In order to be able to compare the magnitude of different multipole tensors, a normalization independent quantity has been introduced, the polarization 
\begin{align} 
\pi^{kpr}_{t}=2(2\ell+1)(2k+1)(2p+1)(2r+1) |N_{kpr\ell}\,{w}^{kpr}_{t}|^2\,.
\end{align}
All contributions, excluding $kpr=000$,
add up to a total polarization 
\begin{align} 
\pi^\mathrm{tot}=\sum_{kprt}\pi^{kpr}_{t}\,,
\end{align} 
which is constrained by the inequality 
\begin{align} 
\pi^\mathrm{tot}\leq (10-n_d)n_d\,,\label{ineq}
\end{align} 
where $n_d$ is the occupation number of the $d$-shell.

In Table \ref{pol} the results for the full TR even calculation are presented in terms of the largest contributions to the 
exchange energy as well as the polarization. 
For the $j_\mathrm{eff}$-model we notice that sum of three non-zero contributions is always 20, but the individual contributions depend on the parameter $a$ of \eqref{ham}. This can be easily understood from the fact that for a dominating CF or SOC term the corresponding multipole tensor polarisation, $\pi^{404}$ and $\pi^{110}$ respectively, takes the largest values.
For the full calculations we notice that while the same three polarizations are largest they add up to 3.5 rather than 20. This is a signature that the $j_\mathrm{eff}$-model is not perfectly valid, which is due to the distorted and rotated oxygen octahedras surrounding the Ir site.

\subsection{TR symmetry breaking}
\label{TR-symm-Break}

As discussed above under Secion \ref{CF-SOC} the degenerate levels of Eq.~\eqref{phi-deg} are split with the addition of the TR odd contribution $\cH_U$.
The details of this splitting depend on the degeneracy parameters $\beta$ and $\delta$ and which multipole component is mainly responsible. This will lead to varying observables in the broken symmetry solutions. For instance the spin moment direction is intimately connected to the value of $\beta$ and $\delta$ as can be seen in Fig.~1 of the main article.

\begin{figure}[htbp]
\begin{center}
\includegraphics[width=0.7\columnwidth]{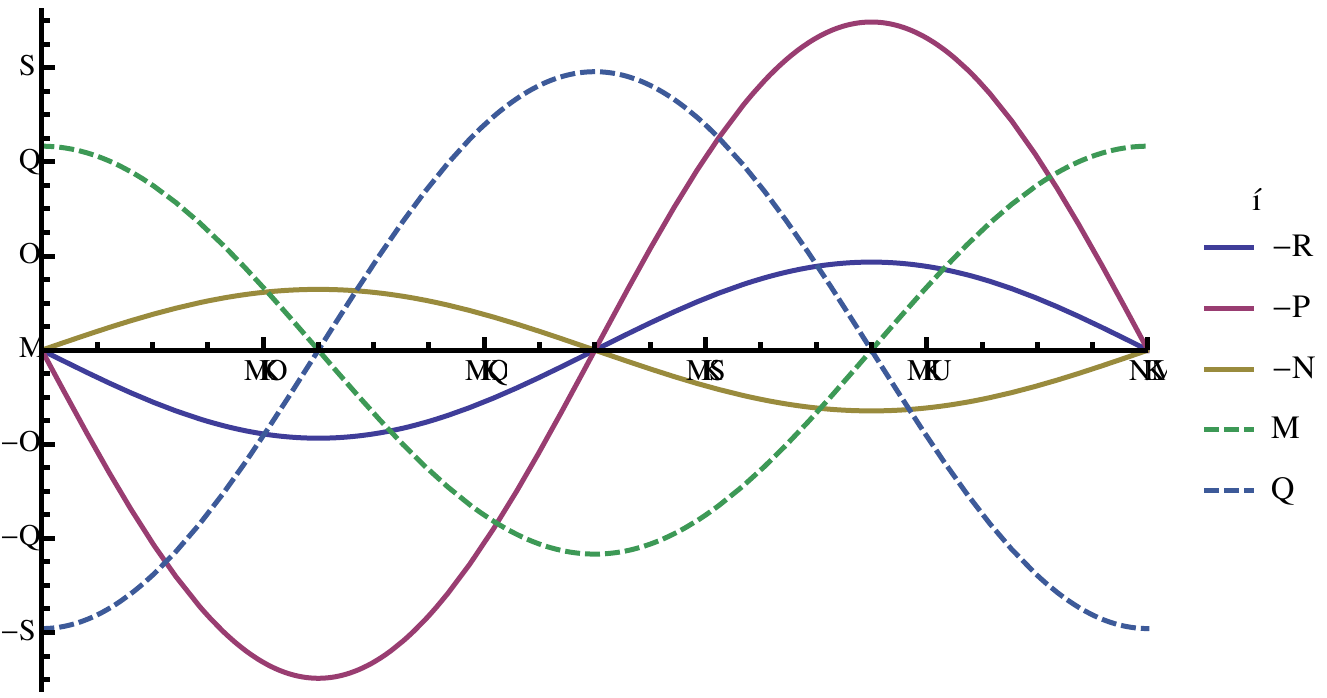}
\caption{ }
\label{MP}
\end{center}
\end{figure}

In the most simple version of $j_\mathrm{eff}$-model the $e_{5/2g}$ states are localized and discrete,
which imply that many TR odd tensor multipoles (odd $k+p$) will have non-vanishing expectation values. For instance
in Fig.~\ref{MP}, we have plotted
\begin{align} 
w^{415}_t(\beta)=\psi_2^\dagger(\beta,\pi/2)\,\Gamma^{415}_t\,\psi_2^{\phantom{\dagger}}(\beta,\pi/2)\,,
\end{align}
for the non-vanishing components $t$ of the TR odd triakontadipoles with $k=4$, $p=1$ and $r=5$, as a function of $\beta/\pi$. However, the corresponding polarization $\pi^{415}$ does not depend on the angles.
The fact that the states are localized also leads to that the inequality of Eq.~\eqref{ineq} becomes an equality, which is confirmed by the polarizations of Table  \ref{pol}, as the TR odd polarisation add up to 5 and the TR even is 20 while $n_d$=5.

In the full calculation the total polarization is much smaller which is a sign that the involved states are more band-like in nature.
From Table \ref{pol} we notice that the largest TR symmetry breaking polarization comes from $\pi^{415}$ and the next largest from $\pi^{213}$,
the same that dominate the $j_\mathrm{eff}$-model but smaller with approximately a factor 5. The ordinary spin polarization $\pi^{011}$ is however negligible and even smaller by a factor 15 than the already small value of the $j_\mathrm{eff}$-model.

\begin{table}
\centering 
\caption{}
\begin{tabular} {|l|l|c|c|c|}
\hline  
TR& $kpr$ & $E_\mathrm{X}^{kpr}$ (meV) & $\pi^{kpr}$ & $\pi_{j_\mathrm{eff}}$\\
\hline
&110 & $-158.5$ & 1.134 & $20^*$\\
even&111 & $-1.0$ & 0.007 & 0\\
&314 & $-16.1$ & 0.184 & $20^*$\\
&404 & $-433.5$ & 2.426 & $20^*$\\
\hline
&011 & $-2.4$ & 0.007 & 0.111\\
&101 & $-4.2$ & 0.030 & 0.222\\
&211 & $-1.6$ & 0.010 & 0.063\\
odd&213 & $-41.3$ & 0.268 & 1.524\\
&303 & $-8.5$ & 0.098 & 0.889\\
&414 & $-1.5$ & 0.008 & 0\\
&415 & $-71.8$ & 0.402 & 2.134 \\
\hline 
\hline
&total&& 4.585 & 25 \\
\hline
\end{tabular}
\label{pol}
\end{table}

The TR breaking due to spontaneous formation of an order parameter  OP in terms of multipole tensor components can be understood as follows. 
We consider the largest component of the dominant polarization, which is a slightly rotated $w^{415}_{-3}$ multipole, as the primary OP, i.e.~which is main responsible to break the TR symmetry.

In the calculation there is a linear combination of tensor components $w^{415}_{\pm3}$ which take the largest values and they can be viewed as a rotation of the largest component, $w^{415}_{-3}$ by an angle $\theta$ around the $z$-axis. 
Then the appearance of these TR odd multipole tensors in the ground state gives rise to splitting of the TR even solution by the auxiliary field of Eq.~\eqref{hamu} which is  a matrix in the local basis 
and proportional to the magnitude of the rotated tensor moment $\tilde{w}^{415}_{-3}$  
\begin{align}
\cH_U\approx 2  K_{415}\, \tilde{w}^{415}_{-3}\, \tilde{\Gamma}^{415}_{-3}(\theta)\,,\label{field}
\end{align}
where $K_{415}$ can be obtained through Eqs.~\eqref{K_kpr} and \eqref{A_k} as a linear combination of the three Racah  parameters \cite{Polarisation}
\begin{align} 
K_{415}=-\frac{E^{(0)}+2E^{(1)}+5E^{(2)}}{504}\,.
\end{align} 

In this case the operator of Eq.~\eqref{field} takes the matrix form
\begin{align} 
\tilde{\Gamma}^{415}_{-3}
&=
e^{-i\theta J_{z}}{\Gamma}^{415}_{-3}e^{i\theta J_{z}}=\cos 3\theta\, \Gamma^{415}_{-3}-\sin 3\theta\, \Gamma^{415}_{3}=   \nonumber \\
&\left(
\begin{array}{cccccccccc}
 0 & 0 & 0 & 0 & 0 & 0 & 0 & 0 & 0 & 0 \\
 0 & 0 & 0 & 0 & 0 & 0 & 0 & 0 & 0 & 0 \\
 0 & 0 & 0 & 0 & 0 & 0 & 0 & 0 & 0 & 0 \\
 0 & 0 & 0 & 0 & 0 & 0 & 0 & 0 & 0 & 0 \\
 0 & 0 & 0 & 0 & 0 & 0 & 0 & -2 i \sqrt{7} e^{3 i \theta } & 0 & 0 \\
 0 & 0 & 0 & 0 & 0 & 0 & 0 & 0 & i \sqrt{70} e^{3 i \theta } & 0 \\
 0 & 0 & 0 & 0 & 0 & 0 & 0 & 0 & 0 & -2 i \sqrt{7} e^{3 i \theta } \\
 0 & 0 & 0 & 0 & 2 i \sqrt{7} e^{-3 i \theta } & 0 & 0 & 0 & 0 & 0 \\
 0 & 0 & 0 & 0 & 0 & -i \sqrt{70} e^{-3 i \theta } & 0 & 0 & 0 & 0 \\
 0 & 0 & 0 & 0 & 0 & 0 & 2 i \sqrt{7} e^{-3 i \theta } & 0 & 0 & 0 \\
\end{array}\label{Gamma-mat}
\right)
\end{align}
The two largest eigenvalues (in magnitude) of this rotated operator are $\lambda_{1,2}=\pm\sqrt{70}$  and the corresponding eigenvectors are
\begin{align} 
 \phi_1&=\frac{1}{\sqrt{2}}\left\{ \left|\frac{5}{2},\frac{3}{2}\right>+i e^{3i\theta}\left|\frac{5}{2},-\frac{3}{2}\right>\right\}\nonumber\\
\phi_2&=\frac{1}{\sqrt{2}}\left\{ \left|\frac{5}{2},\frac{3}{2}\right>-i e^{3i\theta}\left|\frac{5}{2},-\frac{3}{2}\right>\right\}\,,\label{eigvec}
\end{align}
respectively.

Then from the eigenvectors of Eq.~\eqref{eigvec} we can readily see that the presence of the OP $\tilde{w}^{415}_{-3}$ primarily splits the degenerate $j=5/2$, $m_j=\pm3/2$ 
states, that  dominate around the Fermi energy for the TR even case through the action of Eq.~\eqref{field}. 

This was illustrated by the DOS projected upon these the eigen-vectors $\phi_1$ and $\phi_2$ that were displayed in Fig.~3 of the main paper. 

The support  from 
the Swedish Research Council (VR) is thankfully acknowledged. 
The calculations have 
been performed at the Swedish high performance centers 
HPC2N and NSC 
under grants provided by the 
Swedish National Infrastructure for Computing (SNIC).

\end{document}